\documentclass[conference]{IEEEtran}
\IEEEoverridecommandlockouts
\usepackage{cite}
\usepackage{amsmath,amssymb,amsfonts}
\usepackage{algorithmic}
\usepackage{graphicx}
\usepackage{textcomp}
\usepackage{xcolor}
\usepackage{booktabs}
\usepackage[font=small,labelfont=bf]{caption}
\usepackage{listings}
\usepackage{placeins}
\usepackage[utf8]{inputenc}
\usepackage[hidelinks]{hyperref}
\def\BibTeX{{\rm B\kern-.05em{\sc i\kern-.025em b}\kern-.08em
    T\kern-.1667em\lower.7ex\hbox{E}\kern-.125emX}}

\begin{document}

\title{Hand-Written PTX Tensor-Core GEMM Kernels:\\A Multi-Precision Study on NVIDIA L4}
\author{
\IEEEauthorblockN{Matt J. Borowski, MSc}
\IEEEauthorblockA{Kernel Engineer \\
ORCID: \href{https://orcid.org/0009-0005-6888-8020}{0009-0005-6888-8020}}
\and
\IEEEauthorblockN{Blazej Osinski, PhD}
\IEEEauthorblockA{Machine Learning Engineer \\
ORCID: \href{https://orcid.org/0000-0003-1821-7796}{0000-0003-1821-7796}}
}

\maketitle

\begin{abstract}
High-performance Tensor Core kernels rely on a low-level PTX pipeline built from asynchronous data movement with \texttt{cp.async}, warp-level matrix loads with \texttt{ldmatrix}, and matrix multiply-accumulate operations with \texttt{mma.sync}. However, most application code accesses Tensor Cores indirectly through the WMMA C++ API. This paper asks a focused, practical question: when does replacing WMMA with hand-written PTX actually pay off?
To answer this question, we conduct a controlled, single-GPU study on an NVIDIA L4 GPU (Ada, SM89), comparing double-buffered WMMA baselines with a family of hand-written PTX GEMM kernels across FP16, INT8, and INT4 arithmetic and square problem sizes from $N=512$ to $N=8192$.
Every kernel is profiled with Nsight Compute across the full metric set, and PTX speedups are reported relative to the corresponding same-precision WMMA baseline.
Hand-written PTX provides no end-to-end speedup for FP16, because its instruction-level gains are offset by operand-packing overhead.
In contrast, the PTX kernels achieve consistent speedups of $1.4\times$--$1.8\times$ for INT8, driven primarily by lower instruction counts and better global-memory coalescing, and $2.9\times$--$4.3\times$ for INT4, where native \texttt{mma.sync.m16n8k64.s4} execution avoids the software-emulated sequence used by the WMMA path.
Relative to the FP16 WMMA baseline, the best quantized kernels reach $34.4\times$ (INT8) and $98.7\times$ (INT4) at $N=8192$.
Across these experiments, occupancy is a poor predictor of throughput. For large matrices, performance instead tracks memory-system behavior---particularly global-load coalescing and DRAM-active cycles---more closely than Tensor Core utilization. These results identify the precisions and operating regimes in which the additional complexity of hand-written PTX is justified.
\end{abstract}

Index Terms---GPU computing, Tensor Cores, PTX, GEMM, quantization, INT4, INT8, \texttt{mma.sync}, \texttt{cp.async}, Nsight Compute, NVIDIA Ada, LLM inference.

\section{Introduction}

General matrix multiplication (GEMM) dominates the compute cost of modern deep-learning inference, and on NVIDIA GPUs the fastest path for GEMM is the Tensor Core. Since the Volta generation, Tensor Cores have been exposed to developers at two very different levels of abstraction. The high-level WMMA (warp-matrix-multiply-accumulate) C++ API is convenient and portable, but it fixes fragment shapes, load patterns, and accumulator layouts, and it hides the underlying instruction stream. The low-level PTX path---\texttt{cp.async} for asynchronous global-to-shared copies, \texttt{ldmatrix} for cooperative shared-to-register fragment loads, and \texttt{mma.sync} for the matrix multiply itself---gives the programmer direct control over tile shape, staging depth, and register mapping, at the cost of considerable complexity.

A widely held assumption is that hand-written PTX is uniformly faster than the WMMA API. In practice the picture is far more nuanced: whether PTX helps depends strongly on the numeric precision, the problem size, and, above all, on whether the kernel is compute-bound or memory-bound. This paper provides a controlled, reproducible answer for one concrete platform---the NVIDIA L4 inference GPU (Ada, compute capability 8.9)---by profiling a coherent family of kernels under identical conditions.

We make the following contributions. First, we implement a matched set of PTX GEMM kernels across FP16, INT8, and INT4 that share a common tiling strategy and differ only in the instruction-level choices under study (Section~\ref{subs:kerneldesignspace}). Second, we benchmark each kernel against a double-buffered WMMA baseline of the same precision, over N=512 to N=8192, using Nsight Compute for full hardware-counter attribution (Section~\ref{section:results}). Third, we isolate the mechanisms behind each result: instruction count and coalescing for INT8, avoidance of WMMA INT4 software emulation for INT4, and the compute-to-memory-bound transition that flattens all FP16 differences at large N. Finally, we perform a focused ablation over the best INT4 kernel family---loader split, cache-eviction policy, and operand layout---to show that the chosen configuration sits in a genuine local optimum (see Section~\ref{subs:int4k64}). As part of these contributions, the complete source code of every kernel studied here---together with the build system, benchmarking harness, and all Nsight Compute profiles---is openly available as the article's reproducibility artifact (see the artifact availability statement at the end of Section~\ref{section:conclusion}).

Taken together, the results yield a practical rule of thumb: PTX is worth the engineering effort precisely when it removes instruction overhead that the WMMA path cannot avoid---most dramatically for INT4, where WMMA emulates the operation in software---and is not worth it when the kernel is already bandwidth-bound or when the compiler's WMMA lowering is already near-optimal, as with FP16. These findings are directly relevant to serving quantized open-weight LLMs on a single commodity inference GPU.

\section{Background}

\subsection{Tensor Cores, WMMA, and PTX \texttt{mma.sync}}

A Tensor Core performs a small matrix multiply-accumulate,
\(D = AB + C\), through a single warp-collective instruction.
The WMMA API exposes this operation through opaque fragment types and the \texttt{load\_matrix\_sync}, \texttt{mma\_sync}, and \texttt{store\_matrix\_sync} calls, using the \texttt{m16n16k16} shape for FP16 and INT8.
The PTX layer instead exposes the underlying \texttt{mma.sync.aligned} instructions with explicit shapes such as \texttt{m16n8k16} and \texttt{m16n8k64}, together with \texttt{ldmatrix} for loading the register fragments held by each lane.
Working at the PTX level lets a kernel choose the \(K\)-tile width, decompose one WMMA step into several narrower MMA instructions, and control exactly how many instructions are issued per unit of arithmetic work.

\subsection{Asynchronous staging: \texttt{cp.async} and pipelining}

All kernels studied here overlap DRAM$\to$shared-memory prefetch with tile-level compute. The \texttt{cp.async} instruction copies a 16-byte vector from global to shared memory without occupying registers or blocking the issuing warp; \texttt{commit\_group} and \texttt{wait\_group} bracket the in-flight copies. A two-stage (double-buffered) pipeline keeps one shared-memory buffer filling while the other is consumed by \texttt{ldmatrix} + \texttt{mma.sync}; a three-stage variant (\texttt{wait\_group} 1) keeps an additional buffer in flight to hide longer L2 latency at the cost of extra shared-memory pressure.

\subsection{Quantization and the memory wall}

Lowering operand precision from FP16 to INT8 or INT4 shrinks the bytes moved per multiply-accumulate by 2$\times$ and 4$\times$ respectively, and shrinks the working set proportionally. For inference-scale GEMM this is decisive: once operands are narrow enough for the active tiles to reside in L2, the kernel escapes the DRAM bottleneck that dominates FP16 at large N. The experiments below quantify this directly and show that quantization reduces memory-bandwidth consumption by roughly 5$\times$--100$\times$ while maintaining or improving cache locality.

\subsection{Platform: NVIDIA L4 (Ada, SM89)}

All measurements were taken on an NVIDIA L4, an Ada-generation datacenter inference GPU (compute capability 8.9) with fourth-generation Tensor Cores supporting FP16, INT8, and INT4 MMA. The optimization patterns transfer to other Tensor-Core GPUs such as A100 (SM80) and H100 (SM90), though the exact gains are hardware- and compiler-dependent; only the SM target passed to the compiler changes.

\section{Methodology}

\subsection{Kernel design space}
\label{subs:kerneldesignspace}

Every kernel implements a tiled GEMM with the warp- and block-level tiling parameters fixed by a shared configuration header (\texttt{WMMA\_M} = \texttt{WMMA\_N} = 16, four warp-tiles in X and two in Y, eight warps per block). Within each precision family, exactly one WMMA-API kernel serves as the baseline and the PTX variants change one dimension of the design at a time: the SRAM$\to$register load path (\texttt{ldmatrix} width vs. manual scalar packing), the \texttt{mma.sync} tile shape (K = 8/16/32/64), the accumulator type, and the pipeline depth (two- vs. three-stage). This one-variable-at-a-time discipline is what allows each observed speedup or slowdown to be attributed to a specific mechanism.

Listing~\ref{lst:int4-double-buffer} shows the core of the fastest large-N INT4 kernel (\texttt{int4\_ptx\_mma\_k64}): a double-buffered main loop that issues \texttt{cp.async} prefetches for the next K-tile, performs two \texttt{m16n8k64} INT4 MMA calls on the current fragments, then rotates buffers. Listing~\ref{lst:int4-primitives} isolates the three PTX primitives---\texttt{ldmatrix.x4}, \texttt{ldmatrix.x2}, and \texttt{mma.sync.m16n8k64.s4}---that distinguish the native path from WMMA. Listing~\ref{lst:int4-three-stage} shows how the three-stage variant keeps one additional copy group in flight with \texttt{wait\_group} 1.

\begin{lstlisting}[language=C++,float=*,floatplacement=t,
caption={Overlap of \texttt{cp.async} prefetch with \texttt{mma.sync} compute in the double-buffered INT4 k64 kernel.},
label={lst:int4-double-buffer}]
// int4_ptx_mma_k64 -- double-buffered main loop (K-tile = 64 int4)
for (int k = WMMA_K; k < K; k += WMMA_K) {
    const int next = 1 - buf;

    // Stage next A/B tile: 16-byte cp.async vectors, .ca cache policy
    for (int i = lane_id; i < (WMMA_M * K_BYTES) / 16;
         i += THREADS_PER_WARP) {
        const int row = (i * 16) / K_BYTES;
        const int byte_col = (i * 16) % K_BYTES;
        const unsigned dst = __cvta_generic_to_shared(
            &As[next][warp_id][row][byte_col]);
        asm volatile(
            "cp.async.ca.shared.global [%0], [%1], 16;"
            :: "r"(dst),
               "l"(&A_b[(tile_row + row) * (K / 2) + byte_col + (k / 2)]));
    }
    // ... identical staging for B ...
    asm volatile("cp.async.commit_group;");

    mma_int4_k64(rc0, ra, rb0);  // compute on current fragments
    mma_int4_k64(rc1, ra, rb1);

    asm volatile("cp.async.wait_group 0;");
    __syncthreads();
    buf = next;  // rotate double buffer

    ldmatrix_a_k64(ra, As[buf][warp_id], lane_id);
    ldmatrix_b_k64(rb0, Bs[buf][warp_id], lane_id, 0);
    ldmatrix_b_k64(rb1, Bs[buf][warp_id], lane_id, 8);
}
\end{lstlisting}

\begin{lstlisting}[language=C++,float=*,floatplacement=t,
caption={The three PTX primitives that define the native INT4 path: cooperative \texttt{ldmatrix} fragment loads and the \texttt{m16n8k64.s4} MMA.},
label={lst:int4-primitives}]
// (a) A fragment: ldmatrix.x4 loads a 16x64 int4 tile (as 16x16 b16)
asm volatile(
    "ldmatrix.sync.aligned.m8n8.x4.shared.b16 "
    "{%0,%1,%2,%3}, [%4];"
    : "=r"(t0), "=r"(t1), "=r"(t2), "=r"(t3)
    : "r"(addr));

// (b) B fragment: ldmatrix.x2 per n8-half (8x64 int4 tile)
asm volatile(
    "ldmatrix.sync.aligned.m8n8.x2.shared.b16 {%0,%1}, [%2];"
    : "=r"(rb[0]), "=r"(rb[1])
    : "r"(addr));

// (c) Native INT4 Tensor-Core multiply-accumulate (no WMMA emulation)
asm volatile(
    "mma.sync.aligned.m16n8k64.row.col.s32.s4.s4.s32 "
    "{%0,%1,%2,%3}, {%4,%5,%6,%7}, {%8,%9}, {%0,%1,%2,%3};"
    : "+r"(c0), "+r"(c1), "+r"(c2), "+r"(c3)
    : "r"(ra[0]), "r"(ra[1]), "r"(ra[2]), "r"(ra[3]),
      "r"(rb[0]), "r"(rb[1]));
\end{lstlisting}

\subsection{Benchmark and profiling setup}

Each kernel is compiled for \texttt{CUDA\_ARCH}=89 and run over square sizes N $\in$ \{512, 1024, 2048, 4096, 8192\}. Wall-clock timings are collected from repeated launches; detailed hardware counters are collected with Nsight Compute (ncu --set full), which reports throughput, occupancy, cache hit rates, instruction counts, coalescing efficiency, and warp-state statistics. Speedup is always defined relative to the same-precision WMMA baseline (\texttt{fp16\_wmma}, \texttt{int8\_wmma}, \texttt{int4\_wmma}); a separate summary also expresses every kernel relative to the FP16 WMMA baseline to capture the end-to-end value of quantization plus PTX.

The study is organized as four runs: Run 1 covers the FP16 family, Run 2 the INT8 family, Run 3 the INT4 base kernels, and Run 4 an ablation over the best INT4 (k64) family. Table~\ref{tab:run-summary} summarizes the outcome of each run before the detailed analysis.

\begin{table*}[t]
\renewcommand{\arraystretch}{1.3}
\caption{AT-A-GLANCE RESULTS: BEST KERNEL AND DOMINANT MECHANISM PER RUN}
\label{tab:run-summary}
\centering
\normalsize
\begin{tabular}{p{0.18\linewidth}p{0.18\linewidth}p{0.18\linewidth}p{0.28\linewidth}}
\hline
Run & Best kernel(s) & Speedup vs. same-precision WMMA & Key finding \\
\hline
Run 1 $\cdot$ FP16 & \texttt{fp16\_wmma} & 1.0$\times$ & PTX does not improve wall time; local gains are offset by extra instruction / packing overhead. \\
Run 2 $\cdot$ INT8 & \texttt{int8\_ptx\_mma\_k32} & 1.4$\times$--1.8$\times$ & k32 wins via fewer executed instructions and near-perfect global-load coalescing. \\
Run 3 $\cdot$ INT4 & \texttt{int4\_ptx\_3stage} (small N); \texttt{int4\_ptx\_mma\_k64} (large N) & 2.9$\times$--4.3$\times$ & Both avoid WMMA INT4 software emulation; 3stage loses L1 locality as N grows, k64 keeps it. \\
Run 4 $\cdot$ INT4 k64 & \texttt{int4\_ptx\_mma\_k64}\_ \texttt{x4\_x2nontrans\_ca} & 2.9$\times$--4.3$\times$ & Loader split, cache policy and layout do not beat the baseline; transposed B collapses coalescing. \\
\hline
\end{tabular}
\end{table*}

\begin{table*}[t]
\renewcommand{\arraystretch}{1.3}
\caption{HEADLINE PER-KERNEL METRICS AT N = 8192}
\label{tab:headline-metrics}
\centering
\normalsize
\begin{tabular}{p{0.18\linewidth}p{0.04\linewidth}p{0.12\linewidth}p{0.08\linewidth}p{0.06\linewidth}p{0.08\linewidth}p{0.08\linewidth}p{0.09\linewidth}}
\hline
Kernel & Prec. & Duration (ms) & AI & TOPS & DRAM \% & L2 hit \% & L1/TEX \% \\
\hline
\texttt{fp16\_wmma} & fp16 & 16500 & 2.73 & 0.067 & 54.0 & 65.9 & 26.7 \\
\texttt{int8\_wmma} & int8 & 858.3 & 5.46 & 1.281 & 52.2 & 81.9 & 33.6 \\
\texttt{int8\_ptx\_mma\_k32} & int8 & 480.1 & 5.46 & 2.290 & 52.5 & 65.6 & 61.5 \\
\texttt{int4\_wmma} & int4 & 713.0 & 10.92 & 1.542 & 0.88 & 99.7 & 80.3 \\
\texttt{int4\_ptx\_3stage} & int4 & 197.6 & 10.92 & 5.564 & 3.16 & 99.8 & 14.5 \\
\texttt{int4\_ptx\_mma\_k64} & int4 & 167.1 & 10.92 & 6.581 & 3.73 & 99.5 & 61.6 \\
\hline
\end{tabular}
\end{table*}

\begin{table*}[t]
\renewcommand{\arraystretch}{1.3}
\caption{BEST-PER-PRECISION WALL-CLOCK DURATION (MS) VS. PROBLEM SIZE N}
\label{tab:best-durations}
\centering
\normalsize
\begin{tabular}{p{0.18\linewidth}p{0.16\linewidth}p{0.08\linewidth}p{0.08\linewidth}p{0.08\linewidth}p{0.09\linewidth}p{0.09\linewidth}}
\hline
Kernel & Role & 512 & 1024 & 2048 & 4096 & 8192 \\
\hline
\texttt{fp16\_wmma} & Base \& Optimal & 0.184 & 1.430 & 11.850 & 102.350 & 16500.000 \\
\texttt{int8\_wmma} & Base & 0.152 & 1.110 & 8.550 & 68.470 & 858.300 \\
\texttt{int8\_ptx\_mma\_k32} & Optimal & 0.109 & 0.726 & 5.410 & 42.000 & 480.060 \\
\texttt{int4\_wmma} & Base & 0.195 & 1.450 & 11.400 & 90.110 & 712.960 \\
\texttt{int4\_ptx\_3stage} & Optimal (small N) & 0.056 & 0.382 & 2.910 & 24.260 & 197.630 \\
\texttt{int4\_ptx\_mma\_k64} & Optimal (large N) & 0.068 & 0.395 & 2.810 & 21.520 & 167.100 \\
\hline
\end{tabular}
\end{table*}

\begin{table*}[t]
\renewcommand{\arraystretch}{1.3}
\caption{SPEEDUP VS. THE FP16 WMMA BASELINE (HIGHER IS FASTER)}
\label{tab:speedup-vs-fp16}
\centering
\normalsize
\begin{tabular}{p{0.19\linewidth}p{0.16\linewidth}p{0.08\linewidth}p{0.08\linewidth}p{0.08\linewidth}p{0.08\linewidth}p{0.08\linewidth}}
\hline
Kernel & Role & 512 & 1024 & 2048 & 4096 & 8192 \\
\hline
\texttt{int8\_wmma} & Base & 1.2$\times$ & 1.3$\times$ & 1.4$\times$ & 1.5$\times$ & 19.2$\times$ \\
\texttt{int8\_ptx\_mma\_k32} & Optimal & 1.7$\times$ & 2.0$\times$ & 2.2$\times$ & 2.4$\times$ & 34.4$\times$ \\
\texttt{int4\_wmma} & Base & 0.9$\times$ & 1.0$\times$ & 1.0$\times$ & 1.1$\times$ & 23.1$\times$ \\
\texttt{int4\_ptx\_3stage} & Optimal (small N) & 3.3$\times$ & 3.7$\times$ & 4.1$\times$ & 4.2$\times$ & 83.5$\times$ \\
\texttt{int4\_ptx\_mma\_k64} & Optimal (large N) & 2.7$\times$ & 3.6$\times$ & 4.2$\times$ & 4.8$\times$ & 98.7$\times$ \\
\hline
\end{tabular}
\end{table*}

\section{Results and Analysis}
\label{section:results}

We report each run in turn. Throughout, positive speedup means faster than the relevant WMMA baseline. Two regimes recur: a compute-bound regime at small-to-medium N where instruction efficiency dominates, and a memory-bound regime at large N where bandwidth and coalescing dominate. The transition between them is sharpest for FP16, whose large working set overflows L2 between N=4096 and N=8192.

\subsection{Run 1 --- FP16: PTX does not beat the compiler}

\begin{figure*}[t]
\centering
\includegraphics[width=\textwidth]{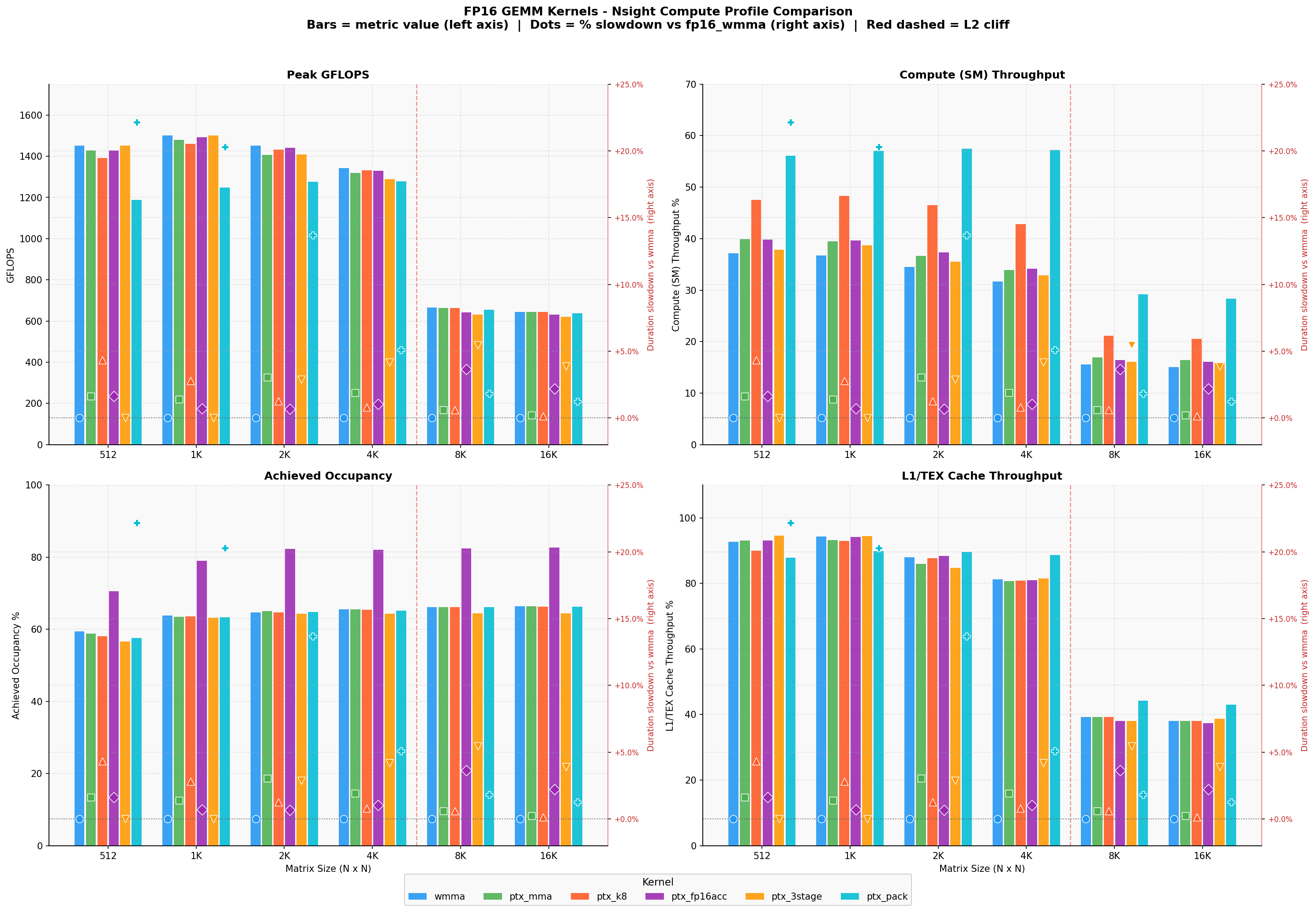}
\caption{FP16 (Run 1): GFLOPS, SM throughput, achieved occupancy, and L1/TEX throughput vs. N. The L2 capacity cliff between N=4096 and N=8192 separates the compute-bound and memory-bound regimes.}
\label{fig:run-1}
\end{figure*}

Six FP16 kernels were profiled. Across all sizes, no hand-written PTX variant is faster than the compiler-optimized WMMA baseline in wall time; measured ratios stay within roughly $\pm$5\% (Table~\ref{tab:fp16-durations}). The reason is visible in the counters: in the compute-bound regime (N $\leq$ 4096) all kernels sustain 1,280--1,500 GFLOPS, and although \texttt{fp16\_ptx\_manual\_pack} reaches the highest SM throughput (\textasciitilde{}57\%) it also carries the most packing overhead, costing up to +22\% at N=512. Once the \textasciitilde{}537 MB combined working set at N=8192 overflows L2, every kernel stalls on DRAM: L1/TEX throughput collapses from \textasciitilde{}90\% to \textasciitilde{}38\%, GFLOPS roughly halve, and all differences shrink below 1\%. The instruction mix becomes irrelevant because bandwidth is the sole bottleneck (Fig.~\ref{fig:run-1}).

A secondary observation concerns the FP16-accumulator variant: using FP16 rather than FP32 accumulators halves the accumulator register count and lifts achieved occupancy to 70--82\% (vs. 56--66\% for the others), yet GFLOPS do not increase. This is direct evidence that occupancy alone does not drive throughput when the kernel is not limited by a warp-count shortage---a theme that recurs in Run 2.

\begin{table*}[t]
\caption{RUN 1 --- FP16 RAW DURATIONS (MS) VS. N}
\label{tab:fp16-durations}
\centering
\normalsize
\begin{tabular}{p{0.24\linewidth}p{0.1\linewidth}p{0.1\linewidth}p{0.1\linewidth}p{0.1\linewidth}p{0.1\linewidth}}
\hline
Kernel & 512 & 1024 & 2048 & 4096 & 8192 \\
\hline
\texttt{fp16\_wmma} (baseline) & 680.155 & 684.000 & 813.093 & 1539.726 & 14098.618 \\
\texttt{fp16\_ptx\_mma} & 710.899 & 698.062 & 782.720 & 1553.984 & 14150.838 \\
\texttt{fp16\_ptx\_k8} & 685.578 & 724.600 & 801.493 & 1540.406 & 14135.938 \\
\texttt{fp16\_ptx\_fp16acc} & 664.341 & 700.616 & 789.057 & 1555.479 & 14553.427 \\
\texttt{fp16\_ptx\_3stage} & 689.617 & 683.957 & 794.762 & 1577.620 & 14828.855 \\
\texttt{fp16\_ptx\_manual\_pack} & 673.823 & 699.585 & 789.764 & 1574.092 & 14312.770 \\
\hline
\end{tabular}
\end{table*}

\begin{table*}[t]
\renewcommand{\arraystretch}{1.3}
\caption{RUN 1 --- FP16 KERNEL DESIGN MATRIX}
\label{tab:fp16-design}
\centering
\normalsize
\begin{tabular}{p{0.18\linewidth}p{0.2\linewidth}p{0.12\linewidth}p{0.04\linewidth}p{0.07\linewidth}p{0.26\linewidth}}
\hline
Kernel & SRAM$\to$Regs & mma shape & Acc & Pipeline & Notes \\
\hline
\texttt{fp16\_wmma} & \texttt{wmma::load\_matrix} \texttt{\_sync} & \texttt{m16n16k16} & f32 & 2-stage & WMMA baseline; no explicit PTX \\
\texttt{fp16\_ptx\_mma} & \texttt{ldmatrix.x4} / \texttt{.x2.trans} & \texttt{m16n8k16} $\times$2 & f32 & 2-stage & first pure-PTX kernel \\
\texttt{fp16\_ptx\_k8} & \texttt{ldmatrix.x2} / \texttt{.x1.trans} & \texttt{m16n8k8} $\times$4 & f32 & 2-stage & narrower K; 4 MMA per K-step \\
\texttt{fp16\_ptx\_fp16acc} & \texttt{ldmatrix.x4} / \texttt{.x2.trans} & \texttt{m16n8k16} $\times$2 & f16 & 2-stage & half the accumulator registers \\
\texttt{fp16\_ptx\_3stage} & \texttt{ldmatrix.x4} / \texttt{.x2.trans} & \texttt{m16n8k16} $\times$2 & f32 & 3-stage & \texttt{wait\_group} 1; extra SRAM buffer \\
\texttt{fp16\_ptx\_manual} \texttt{\_pack} & scalar \texttt{ld.shared} + \texttt{mov.b32} & \texttt{m16n8k16} $\times$2 & f32 & 2-stage & no \texttt{ldmatrix}; exposes packing cost \\
\hline
\end{tabular}
\end{table*}

\subsection{Run 2 --- INT8: k32 wins on instructions and coalescing}

Six INT8 kernels were profiled (Tables~\ref{tab:int8-durations}--\ref{tab:int8-speedup}). \texttt{int8\_ptx\_mma\_k32} is the fastest at every size, from 23\% faster than \texttt{int8\_wmma} at N=512 to 43\% faster at N=8192. Its advantage is rooted in instruction count: by decomposing each K=32 tile step into two tightly unrolled \texttt{m16n8k16} MMA calls it executes 25--43\% fewer instructions than WMMA, and its global-load coalescing is near-perfect---only 0.4\% wasted sectors at N=8192, versus roughly 50\% for every other kernel.

The other variants illuminate the mechanism by contrast. \texttt{int8\_ptx\_mma\_k16} is slower than WMMA at small--medium sizes (uncoalesced loads waste 31 of every 32 bytes per sector) but recovers at N=8192 once a 76\% L1 hit rate absorbs the excess traffic. \texttt{int8\_ptx\_3stage} degrades sharply at N=4096 (+37\%) because its triple-buffer schedule saturates the MIO queue with shared-memory pressure. \texttt{int8\_ptx\_manual\_pack} stays within 2--5\% of WMMA and posts the highest IPC of the group, showing that a dense ALU packing sequence keeps the scheduler fed even without \texttt{ldmatrix}. The scalar \texttt{int8\_dp4a} path is never competitive---up to 4.8$\times$ slower---because \texttt{DP4A} emits several times more instructions per unit of arithmetic and cannot use the Tensor Cores at all.

\begin{figure*}[t]
\centering
\includegraphics[width=\textwidth]{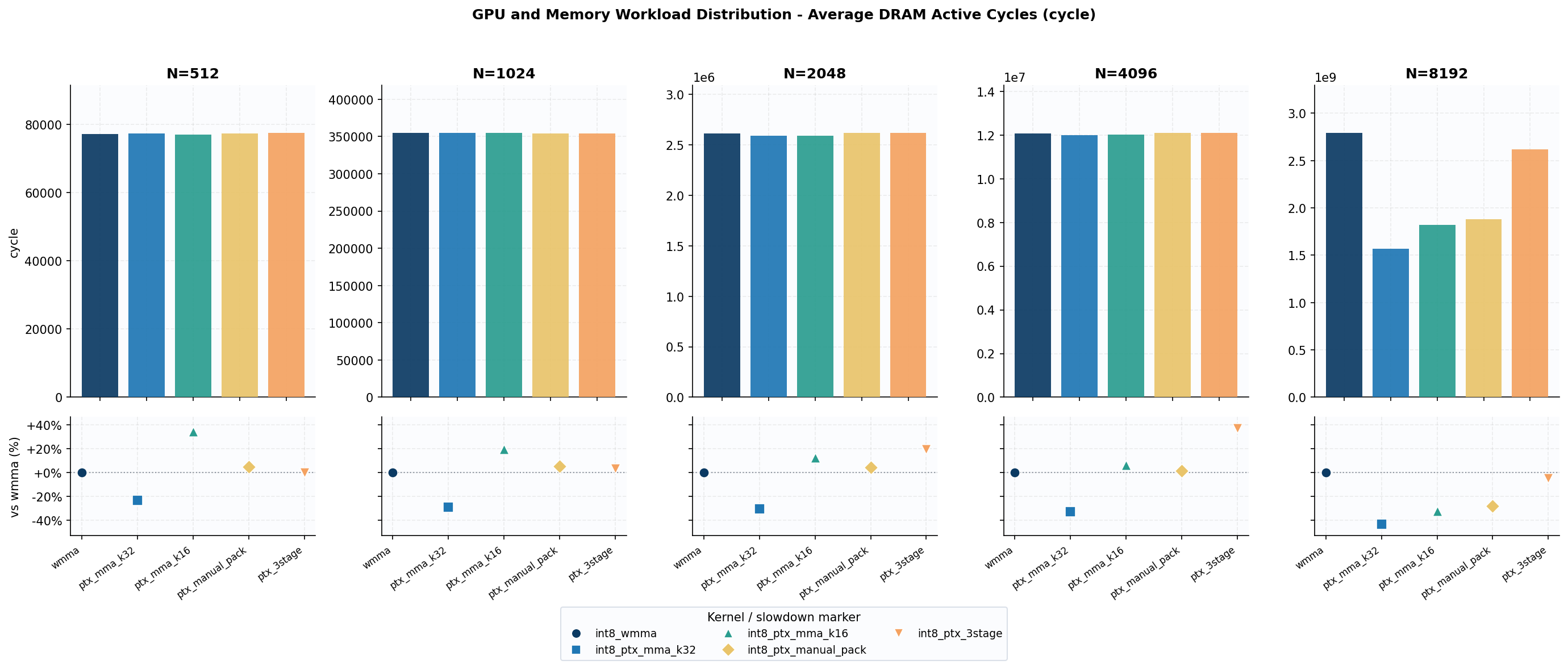}
\caption{INT8 (Run 2): Average DRAM Active Cycles. Flat (compute-bound) for N$\leq$4096, then diverges at N=8192, mirroring the speedup ranking almost 1:1---the clearest evidence that large-N performance is DRAM-latency-bound.}
\label{fig:run-2}
\end{figure*}

Two counter-level findings generalize. First, occupancy is not the performance predictor: \texttt{int8\_wmma} has the highest occupancy at every size yet is consistently beaten by the register-capped k32. Second, at N=8192 the Average DRAM Active Cycles metric is an almost exact fingerprint of wall time---k32 logs 1.57 B DRAM-active cycles versus WMMA's 2.80 B, a $-43.9\%$ reduction that matches its $-43 \% $ wall-clock speedup, and the same correspondence holds for every other kernel (Fig.~\ref{fig:run-2}). This is the clearest evidence that N=8192 is DRAM-latency-bound and that the ranking is set by coalescing quality.

\begin{figure*}[t]
\centering
\includegraphics[width=\textwidth]{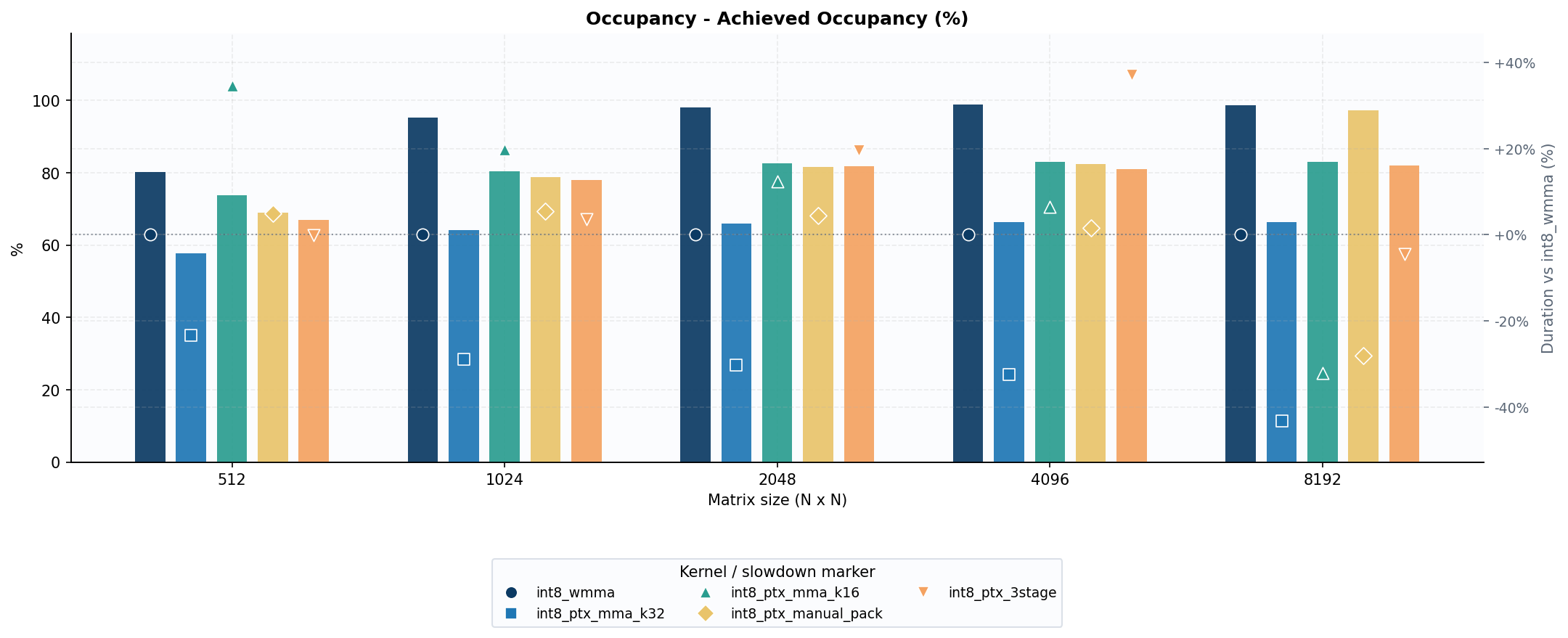}
\caption{INT8 (Run 2): achieved occupancy. The highest-occupancy kernel (\texttt{int8\_wmma}) is not the fastest, confirming that occupancy is not the bottleneck in this regime.}
\label{fig:run-3}
\end{figure*}

\begin{table*}[t]
\renewcommand{\arraystretch}{1.3}
\caption{RUN 2 --- INT8 RAW DURATIONS VS. N}
\label{tab:int8-durations}
\centering
\normalsize
\begin{tabular}{p{0.18\linewidth}p{0.1\linewidth}p{0.1\linewidth}p{0.1\linewidth}p{0.1\linewidth}p{0.1\linewidth}}
\hline
Kernel & 512 & 1024 & 2048 & 4096 & 8192 \\
\hline
\texttt{int8\_wmma} (baseline) & 152.00 µs & 1.11 ms & 8.55 ms & 68.47 ms & 858.30 ms \\
\texttt{int8\_ptx\_mma\_k32} & 108.67 µs & 726.46 µs & 5.41 ms & 42.00 ms & 480.06 ms \\
\texttt{int8\_ptx\_mma\_k16} & 204.54 µs & 1.33 ms & 9.61 ms & 72.93 ms & 582.00 ms \\
\texttt{int8\_ptx\_manual\_pack} & 159.55 µs & 1.17 ms & 8.93 ms & 69.54 ms & 617.68 ms \\
\texttt{int8\_ptx\_3stage} & 151.87 µs & 1.15 ms & 10.22 ms & 93.88 ms & 820.32 ms \\
\texttt{int8\_dp4a} & 588.80 µs & 4.65 ms & 36.75 ms & 296.86 ms & 2360 ms \\
\hline
\end{tabular}
\end{table*}

\begin{table*}[t]
\renewcommand{\arraystretch}{1.3}
\caption{RUN 2 --- INT8 SPEEDUP VS. \texttt{INT8\_WMMA}}
\label{tab:int8-speedup}
\centering
\normalsize
\begin{tabular}{p{0.24\linewidth}p{0.1\linewidth}p{0.1\linewidth}p{0.1\linewidth}p{0.1\linewidth}p{0.1\linewidth}}
\hline
Kernel & 512 & 1024 & 2048 & 4096 & 8192 \\
\hline
\texttt{int8\_ptx\_mma\_k32} & 1.40$\times$ & 1.53$\times$ & 1.58$\times$ & 1.63$\times$ & 1.79$\times$ \\
\texttt{int8\_ptx\_mma\_k16} & 0.74$\times$ & 0.83$\times$ & 0.89$\times$ & 0.94$\times$ & 1.47$\times$ \\
\texttt{int8\_ptx\_manual\_pack} & 0.95$\times$ & 0.95$\times$ & 0.96$\times$ & 0.98$\times$ & 1.39$\times$ \\
\texttt{int8\_ptx\_3stage} & 1.00$\times$ & 0.97$\times$ & 0.84$\times$ & 0.73$\times$ & 1.05$\times$ \\
\texttt{int8\_dp4a} & 0.26$\times$ & 0.24$\times$ & 0.23$\times$ & 0.23$\times$ & 0.36$\times$ \\
\hline
\end{tabular}
\end{table*}

\subsection{Run 3 --- INT4: native MMA beats WMMA emulation}

\begin{figure}[t]
\centering
\includegraphics[width=\linewidth]{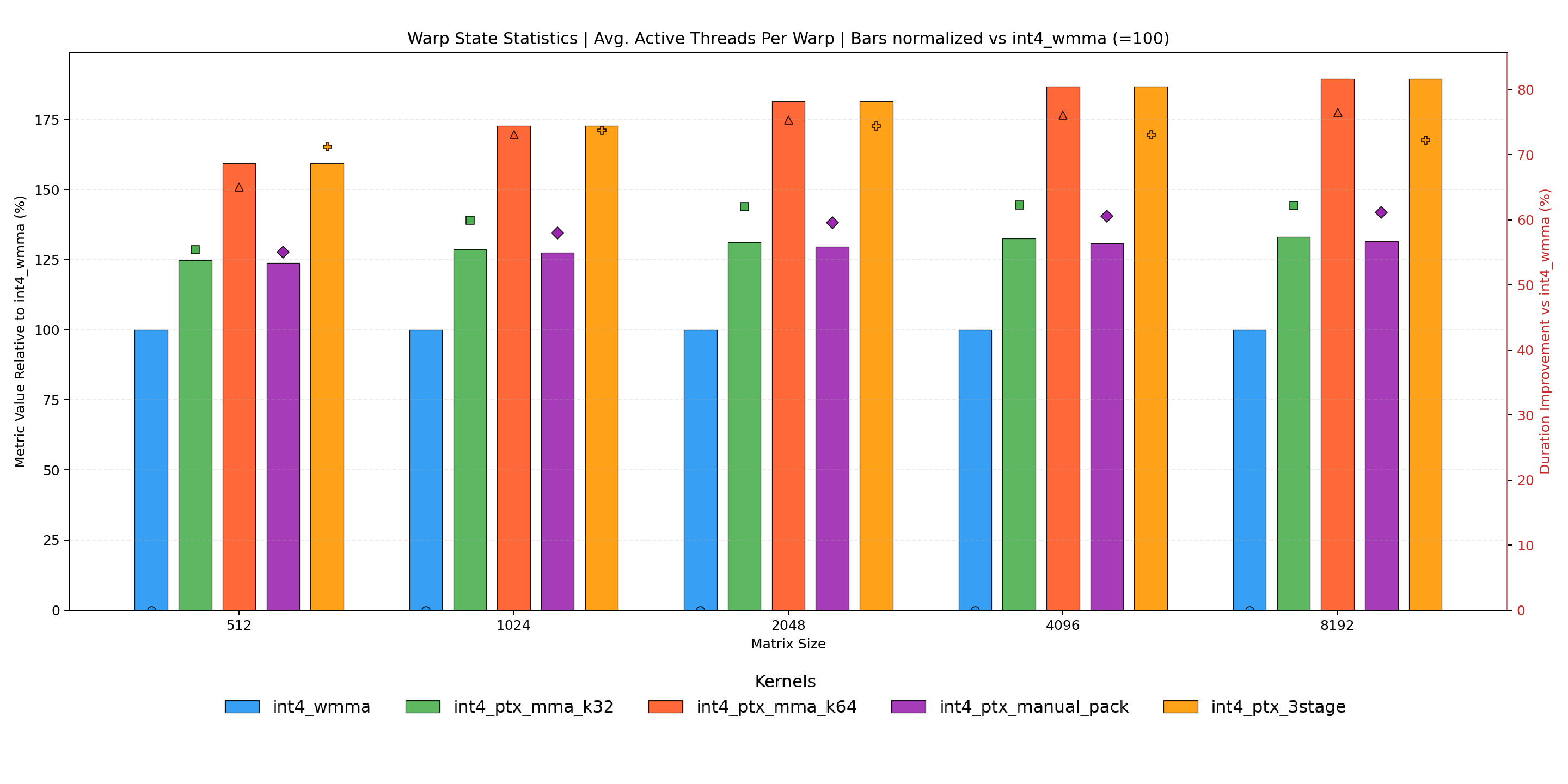}
\caption{INT4 (Run 3): average active threads per warp. The native-MMA kernels hold the full 32/32; \texttt{int4\_wmma} degrades to 16.89 because the WMMA s4 path is software-emulated and lane-divergent.}
\label{fig:run-4}
\end{figure}

The largest gains appear in INT4 (Tables~\ref{tab:int4-durations}--\ref{tab:int4-counters}). All PTX kernels are 2.2$\times$--4.3$\times$ faster than \texttt{int4\_wmma} because the WMMA INT4 path (\texttt{wmma::experimental::precision::s4}) is software-expanded, inflating instruction count and introducing heavy lane-dependent divergence. The native PTX kernels call \texttt{mma.sync.m16n8k64.s4} directly and avoid that expansion entirely. At N=8192, \texttt{int4\_wmma} runs with only 16.89 active threads per warp and 4.6 million divergent branches, while the winning kernels sustain the full 32 active threads per warp with zero divergent branches and execute roughly 6--7$\times$ fewer instructions per scheduler (Fig.~\ref{fig:run-4}).

\begin{figure}[t]
\centering
\includegraphics[width=\linewidth]{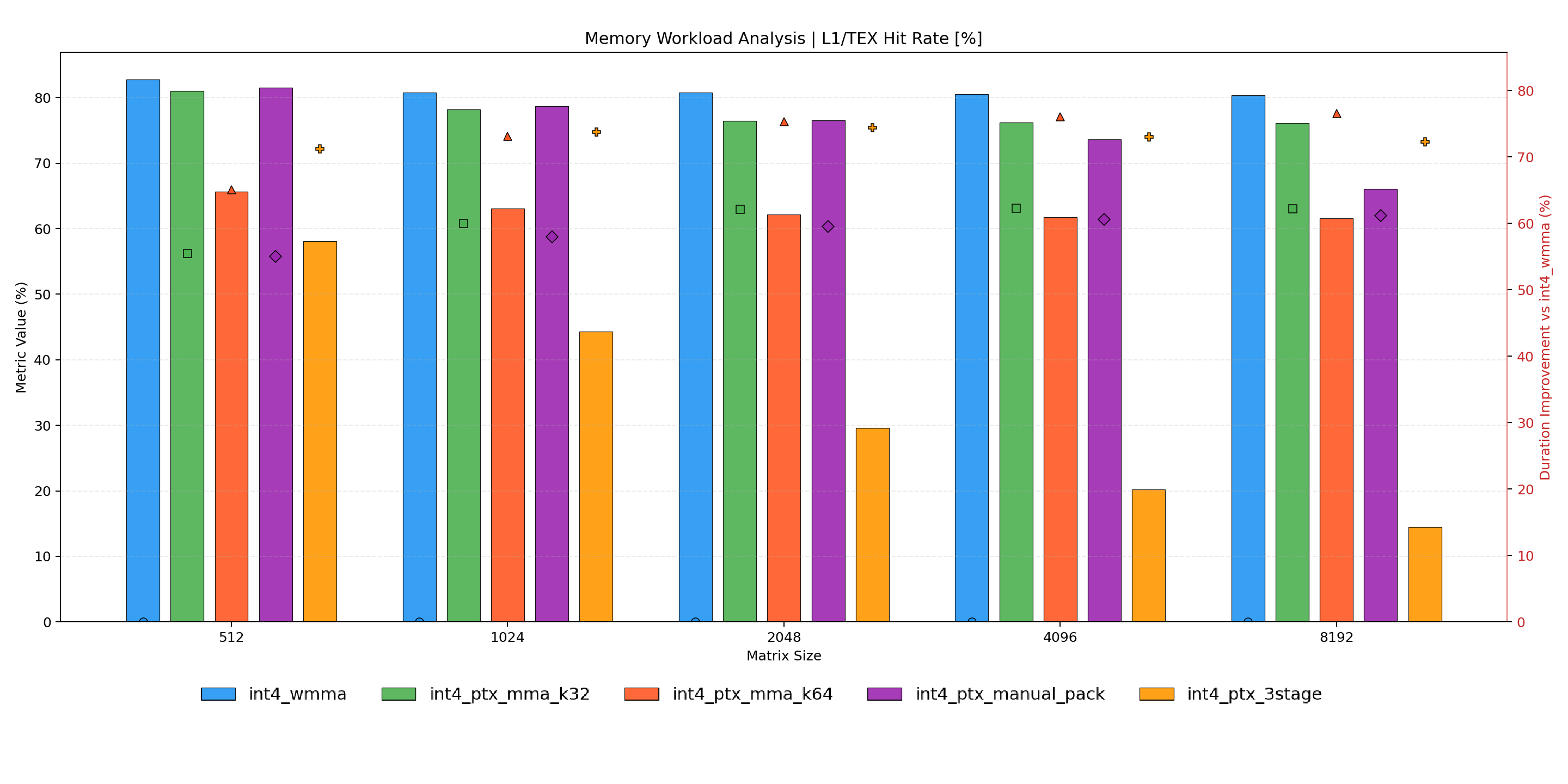}
\caption{INT4 (Run 3): L1/TEX hit rate vs. N. \texttt{int4\_ptx\_mma\_k64} preserves L1 locality (\textasciitilde{}61.6\% at N=8192) where \texttt{int4\_ptx\_3stage} collapses (\textasciitilde{}14.5\%), explaining the large-N crossover.}
\label{fig:run-5}
\end{figure}

Among the PTX kernels the lead varies depending on the GEMM size. \texttt{int4\_ptx\_3stage} is fastest at N=512--1024 thanks to deeper prefetch overlap, but as N grows it becomes strongly L2-driven: by N=8192 its L1/TEX hit rate collapses to \textasciitilde{}14.5\% while it drives L2 throughput to \textasciitilde{}95.6\%. \texttt{int4\_ptx\_mma\_k64} instead retains a \textasciitilde{}61.6\% L1 hit rate at N=8192 and therefore lower large-size latency, making it the best kernel from N=2048 onward (Fig.~\ref{fig:run-5}). The crossover is thus a memory-hierarchy effect, not a difference in tensor-core utilization---both kernels use the identical \texttt{m16n8k64} MMA.

\begin{table*}[t]
\renewcommand{\arraystretch}{1.3}
\caption{RUN 3 --- INT4 RAW DURATIONS VS. N}
\label{tab:int4-durations}
\centering
\normalsize
\begin{tabular}{p{0.24\linewidth}p{0.1\linewidth}p{0.1\linewidth}p{0.1\linewidth}p{0.1\linewidth}p{0.09\linewidth}}
\hline
Kernel & 512 & 1024 & 2048 & 4096 & 8192 \\
\hline
\texttt{int4\_wmma} (baseline) & 195.42 µs & 1.450 ms & 11.40 ms & 90.11 ms & 712.96 ms \\
\texttt{int4\_ptx\_mma\_k32} & 87.01 µs & 0.579 ms & 4.32 ms & 33.99 ms & 269.35 ms \\
\texttt{int4\_ptx\_mma\_k64} & 68.19 µs & 0.395 ms & 2.81 ms & 21.52 ms & 167.10 ms \\
\texttt{int4\_ptx\_manual\_pack} & 87.84 µs & 0.611 ms & 4.61 ms & 35.55 ms & 277.13 ms \\
\texttt{int4\_ptx\_3stage} & 56.16 µs & 0.382 ms & 2.91 ms & 24.26 ms & 197.63 ms \\
\hline
\end{tabular}
\end{table*}

\begin{table*}[t]
\renewcommand{\arraystretch}{1.3}
\caption{RUN 3 --- INT4 SPEEDUP VS. \texttt{INT4\_WMMA}}
\label{tab:int4-speedup}
\centering
\normalsize
\begin{tabular}{p{0.24\linewidth}p{0.1\linewidth}p{0.1\linewidth}p{0.1\linewidth}p{0.1\linewidth}p{0.1\linewidth}}
\hline
Kernel & 512 & 1024 & 2048 & 4096 & 8192 \\
\hline
\texttt{int4\_ptx\_mma\_k32} & 2.25$\times$ & 2.51$\times$ & 2.64$\times$ & 2.65$\times$ & 2.65$\times$ \\
\texttt{int4\_ptx\_mma\_k64} & 2.87$\times$ & 3.68$\times$ & 4.07$\times$ & 4.18$\times$ & 4.27$\times$ \\
\texttt{int4\_ptx\_manual\_pack} & 2.23$\times$ & 2.38$\times$ & 2.48$\times$ & 2.53$\times$ & 2.57$\times$ \\
\texttt{int4\_ptx\_3stage} & 3.48$\times$ & 3.80$\times$ & 3.92$\times$ & 3.72$\times$ & 3.61$\times$ \\
\hline
\end{tabular}
\end{table*}

\begin{table*}[t]
\renewcommand{\arraystretch}{1.3}
\caption{RUN 3 --- KEY COUNTERS AT N = 8192}
\label{tab:int4-counters}
\centering
\normalsize
\begin{tabular}{p{0.28\linewidth}p{0.15\linewidth}p{0.15\linewidth}p{0.15\linewidth}}
\hline
Metric & \texttt{int4\_wmma} & \texttt{int4\_ptx\_mma\_k64} & \texttt{int4\_ptx\_3stage} \\
\hline
Duration (ms) & 712.96 & 167.10 & 197.63 \\
Avg. active threads / warp & 16.89 & 32 & 32 \\
Avg. divergent branches & 4,610,118 & 0 & 0 \\
Avg. executed instr. / scheduler & \textasciitilde{}295 M & \textasciitilde{}45 M & \textasciitilde{}46 M \\
L1/TEX hit rate (\%) & 80.34 & 61.56 & 14.49 \\
L2 throughput (\%) & lower & moderate & 95.6 \\
\hline
\end{tabular}
\end{table*}

\subsection{Run 4 --- INT4 k64 family: a local optimum}
\label{subs:int4k64}

\begin{figure}[t]
\centering
\includegraphics[width=\linewidth]{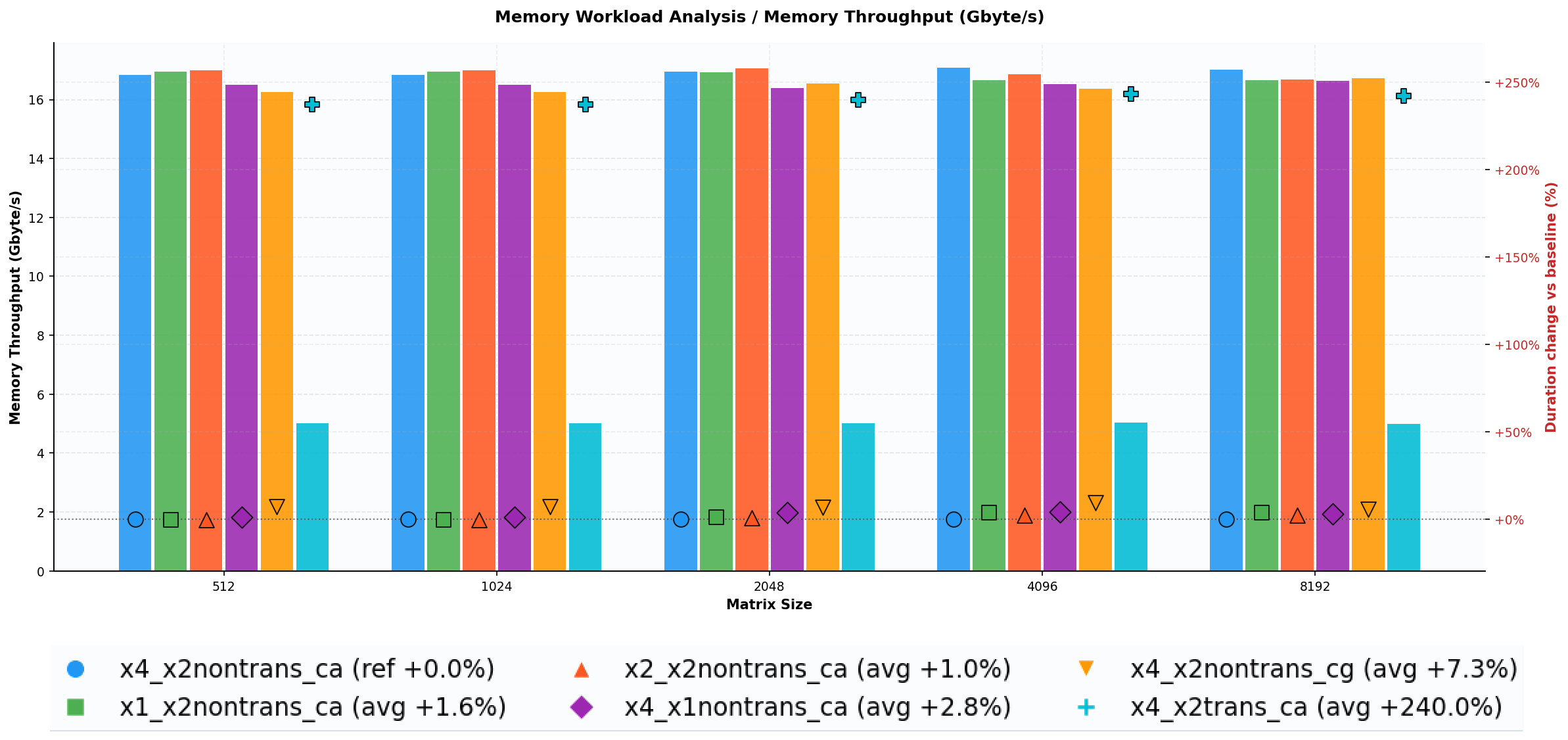}
\caption{INT4 k64 ablation (Run 4): effective memory throughput (GB/s) for different kernels. The transposed-B variant loses most of its usable bandwidth to uncoalesced loads, confirming that coalescing---not MMA arithmetic---governs this kernel family.}
\label{fig:run-6}
\end{figure}

Run 4 fixes the winning \texttt{m16n8k64} MMA strategy and sweeps three knobs: the A-loader split (x1/x2/x4 \texttt{ldmatrix}), the \texttt{cp.async} cache-eviction policy (\texttt{.ca} vs \texttt{.cg}), and the B operand layout (non-transposed vs transposed). Table~\ref{tab:int4-ablation} and Fig.~\ref{fig:run-6} show that the chosen configuration---x4 A-loader, x2 B-loader, non-transposed B, \texttt{.ca} policy---sits in a genuine local-optimum basin. The non-transposed \texttt{.ca} variants differ by only a few percent because they do not change the bottleneck class; they merely redistribute pressure between L1, L2, and instruction issue.

The two off-basin variants fail in instructive ways. The \texttt{.cg} policy pushes traffic from L1 to L2 (L1 hit rate \textasciitilde{}64\% $\to$ \textasciitilde{}20\%, L2 throughput \textasciitilde{}26\% $\to$ \textasciitilde{}50\%), producing a mild but consistent latency regression. The transposed-B variant collapses: with only \textasciitilde{}2.2 of every 32 bytes per global-load sector used, eligible warps per scheduler fall from \textasciitilde{}0.57 to \textasciitilde{}0.20 and warp cycles per instruction more than double, making it up to \textasciitilde{}3.1$\times$ slower at large N. This confirms that coalescing quality, not MMA arithmetic, governs this kernel family.

\begin{table*}[t]
\renewcommand{\arraystretch}{1.3}
\caption{RUN 4 --- INT4 K64 FAMILY SPEEDUP VS. THE \texttt{X4\_X2NONTRANS\_CA} BASELINE}
\label{tab:int4-ablation}
\centering
\normalsize
\begin{tabular}{p{0.18\linewidth}p{0.1\linewidth}p{0.1\linewidth}p{0.1\linewidth}p{0.1\linewidth}p{0.1\linewidth}}
\hline
Variant & 512 & 1024 & 2048 & 4096 & 8192 \\
\hline
\texttt{x1\_x2nontrans\_ca} & 0.97$\times$ & 0.97$\times$ & 0.95$\times$ & 0.94$\times$ & 0.92$\times$ \\
\texttt{x2\_x2nontrans\_ca} & 0.96$\times$ & 0.97$\times$ & 0.96$\times$ & 0.95$\times$ & 0.95$\times$ \\
\texttt{x4\_x1nontrans\_ca} & 0.94$\times$ & 0.95$\times$ & 0.94$\times$ & 0.94$\times$ & 0.93$\times$ \\
\texttt{x4\_x2nontrans\_cg} & 0.94$\times$ & 0.94$\times$ & 0.92$\times$ & 0.90$\times$ & 0.88$\times$ \\
\texttt{x4\_x2trans\_ca} & 0.93$\times$ & 0.84$\times$ & 0.60$\times$ & 0.39$\times$ & 0.32$\times$ \\
\hline
\end{tabular}
\end{table*}

\begin{lstlisting}[language=C++,float=*,floatplacement=t,
caption={Three-stage pipeline using \texttt{wait\_group 1}. Deeper overlap helps at small \(N\) but becomes L2-bound as \(N\) grows.},
label={lst:int4-three-stage}]
// int4_ptx_3stage -- triple-buffered main loop
// (one group kept in flight)
for (int k = 2 * WMMA_K; k < K; k += WMMA_K) {
    const int next = (buf + 1) % 3;
    const int prefetch_buf = (buf + 2) % 3;

    // Prefetch tile two stages ahead
    for (...)
        cp_async16(&As[prefetch_buf][warp_id][row][byte_col], src);
    for (...)
        cp_async16(&Bs[prefetch_buf][warp_id][n][byte_col], src);
    asm volatile("cp.async.commit_group;");

    mma_int4_k64(rc0, ra, rb0);
    mma_int4_k64(rc1, ra, rb1);
    asm volatile("cp.async.wait_group 1;");  // keep 1 group outstanding

    buf = next;
    ldmatrix_a_k64(ra, As[buf][warp_id], lane_id);
    ldmatrix_b_k64(rb0, Bs[buf][warp_id], lane_id, 0);
    ldmatrix_b_k64(rb1, Bs[buf][warp_id], lane_id, 8);
}
\end{lstlisting}

\section{Discussion}

The results support a single organizing principle: PTX is worth its complexity exactly when it removes instruction overhead that the WMMA path cannot avoid. For FP16 the compiler's WMMA lowering is already efficient and the workload becomes bandwidth-bound at scale, so hand-written PTX has nothing to recover and never wins. For INT8 the win is moderate and mechanistic---fewer executed instructions and better coalescing from the k32 decomposition. For INT4 the win is large and structural, because WMMA emulates the s4 operation in software while the PTX kernel issues a single native \texttt{m16n8k64.s4} MMA.

A second principle is that at inference scale the binding constraint is memory, not arithmetic. Quantization helps not only by halving or quartering the bytes per operand but by shrinking the working set enough to stay resident in L2, which is why INT4 escapes the DRAM cliff that caps FP16 at N=8192. The near-1:1 correspondence between Average DRAM Active Cycles and wall time at large N, and the transposed-B collapse in Run 4, both point to coalescing and cache residency as the true levers.

For practitioners serving quantized open-weight LLMs (e.g., Llama-, Mistral-, Qwen-, or Nemotron-class models) on a single L4-class GPU, the practical takeaways are: prefer native INT4/INT8 MMA over WMMA emulation; tune the K-tile and loader split for coalescing before reaching for deeper pipelines; and treat occupancy as a diagnostic, not a target. Because only the SM target passed to the compiler changes, the same kernels are expected to transfer to A100 (SM80) and H100 (SM90), with gains that remain qualitatively similar though quantitatively hardware-dependent.

\section{Scope and Generality}

The present study focuses on a single GPU (NVIDIA L4, SM89), square matrices, and a fixed block/warp tiling; absolute numbers and some rankings may shift on other architectures, non-square shapes, or alternative tilings. Timings for the very largest FP16 case reflect a workload that overflows L2, so small measurement variance is amplified there. Quantized kernels compute INT32 accumulations without in-kernel dequantization or scaling, matching a typical inference pipeline but not end-to-end accuracy; numerical-quality evaluation of the quantized paths is out of scope and left to future work. Finally, the profiling relies on Nsight Compute counter definitions, whose semantics can differ subtly across driver and tool versions.

\section{Conclusion}
\label{section:conclusion}

We presented a controlled, Nsight-Compute-instrumented comparison of hand-written PTX Tensor-Core GEMM kernels against WMMA baselines across FP16, INT8, and INT4 on an NVIDIA L4. PTX offers no benefit for FP16, a 1.4$\times$--1.8$\times$ benefit for INT8, and a 2.9$\times$--4.3$\times$ benefit for INT4 over the same-precision WMMA baseline; relative to FP16 WMMA the best quantized kernels reach 34.4$\times$ (INT8) and 98.7$\times$ (INT4) at N=8192. The gains are explained not by tensor-core utilization but by instruction count, coalescing quality, and cache residency, and occupancy is shown to be a poor predictor of throughput. The work provides both a reusable, precision-matched kernel family and a clear decision rule for when descending to PTX is justified in quantized LLM-inference GEMM.

\vspace{1ex}
\noindent\textbf{Artifact availability.} All kernels, build scripts, Nsight Compute reports, and the profiling charts referenced in this paper are available in the accompanying repository, \url{https://github.com/MattJBorowski1991/TensorCorePTX}, and are permanently archived on Zenodo: \url{https://doi.org/10.5281/zenodo.21815137}.

\FloatBarrier

\end{document}